\documentclass[aps,prl,twocolumn,floatfix,showpacs,superscriptaddress]{revtex4}

\usepackage{graphicx}
\usepackage{amsmath}
\usepackage{amsfonts}
\usepackage{amssymb}
\usepackage[squaren,thinspace]{SIunits}
\usepackage{hyperref}
\usepackage{color}

\begin{document}

\title{Large Tunneling Anisotropic Magnetoresistance mediated by Surface States}

\author{Marie Herv\'e}
\affiliation{Physikalisches Institut, Karlsruhe Institute of Technology,  Wolfgang-Gaede-Stra{\ss}e 1, 76131 Karlsruhe, Germany}

\author{Timofey Balashov}
\affiliation{Physikalisches Institut, Karlsruhe Institute of Technology,  Wolfgang-Gaede-Stra{\ss}e 1, 76131 Karlsruhe, Germany}

\author{Arthur Ernst}
\affiliation{Institute for Theoretical Physics, Johannes Keppler University Linz, Altenberger Stra{\ss}e 69, 4040 Linz, Austria}
\affiliation{Max-Planck-Institut f\"{u}r Mikrostrukturphysik, Weinberg 2, D-06120 Halle, Germany}
  
\author{Wulf Wulfhekel}
\affiliation{Physikalisches Institut, Karlsruhe Institute of Technology,  Wolfgang-Gaede-Stra{\ss}e 1, 76131 Karlsruhe, Germany}

\begin{abstract}
  We investigated the tunneling anisotropic magnetoresistance (TAMR)
  in thick hcp Co films at cryogenic temperatures using scanning
  tunneling microscopy. At around \unit{-350}{\milli\volt}, a strong TAMR
  up to 30\% is found with a characteristic voltage dependence and a
  reversal of sign. With the help of \textit{ab initio} calculations
  the TAMR can be traced back to a spin-polarized occupied surface
  states that experience a strong spin-orbit interaction leading to a
  magnetization direction depending hybridization with bulk states.
\end{abstract}

\date{\today}
\maketitle

Due to its potential for magnetic storage application, the tunneling
anisotropic magnetoresistance (TAMR) effect has attracted a lot of
attention since its discovery
\cite{bode-TAMR,TAMR1,TAMR2,TAMR3,TAMR4,TAMR5}. It is cuased
by changes in the tunneling density of state (DOS) upon magnetization
direction in heterostructures such as
ferromagnet/insulator/normal metal junctions. It originates from the
spin-orbit coupling (SOC), which lifts the degeneracy of electronic states
of a system depending on the magnetization
axis. In the case of an out-of-plane TAMR, the magnetic electrode with
a magnetization in-plane and out-of-plane exhibits two distinct
tunneling DOS \cite{Matos-Abiague2009}. It can be produced by the
Bychkov-Rashba/Dresselhaus SOC field
\cite{Rashba,Dresselhaus,Matos-Abiague2009} or by a change of the
electronic DOS due to SOC induced band splitting \cite{bode-TAMR}. The
TAMR effect was reported for a large diversity of magnetic films. For
example, diluted magnetic semiconductors such as
GaMnAs \cite{TAMR1,TAMR2,TAMR3,TAMR4,TAMR5} display a sizable TAMR. Due to the lack of
inversion symmetry in its Zink-Blend crystalline structure, the
Dresselhaus effect combined with the Bychkov-Rashba effect at the
interfaces produces a TAMR of the order of a few percent. Electronic
states of magnetic 3d metals hybridized with those of 5d transition
metals at interfaces (Fe/W \cite{bode-TAMR}, Co/Pt
\cite{CoPt-exp,CoPt-theo}) constitute a second class of material where
TAMR effects up to 10\% were reported. Recently, a sizable TAMR effect
of 10 \% was reported for a simple fcc Co film epitaxially grown on a
sapphire substrate without the help of 3d/5d interfaces
\cite{hcpCo}. In this case, the uniaxial epitaxial strain induces a
SOC in combination with a Bychkov-Rashba effect and causes a large
TAMR \cite{strainTAMR}.

In order to further increase the TAMR, it was theoretically proposed
to use the enhanced SOC in spin-polarized surface state at metallic
surfaces and interfaces \cite{Chantis,bruno}. Indeed, due to the large
potential gradient, the Rashba effect can increase at surfaces and
interfaces and can strongly affect the electronic band
structure. Depending on the magnetization direction, surface states
can hybridize with bulk states and give rise to surface resonances
that can produce a sizable TAMR. A TAMR of 20\% was theoretically
predicted in a Fe/vacuum/Cu junction \cite{Chantis}. Later a similar
proposition was given for Fe/MgO/Fe magnetic tunnel junctions
\cite{bruno}, where spin-polarized interface resonances are present
\cite{resonanttun1,resonanttun2}. The advantage of this approach is
the strong momentum selectivity of tunneling across the MgO barrier
that can reduce the number of states contributing to tunneling. The
experimental observation, however, revealed a TAMR of only $\approx$1
percent in the Fe/MgO/Fe junctions \cite{IRSTAMR1}.


In this Letter, we report record TAMR effects in hcp Co films of the
order of 30\% without the need of heavy elements. It is caused by
magnetization direction dependent hybridization of spin-polarized
surface states with bulk bands. We argue that this effect is due to
the bulk band structure of Co and as a consequence shows up in many Co
film systems suitable for application.


\begin{figure}[th]
\begin{center}
  \includegraphics[width=8.5cm]{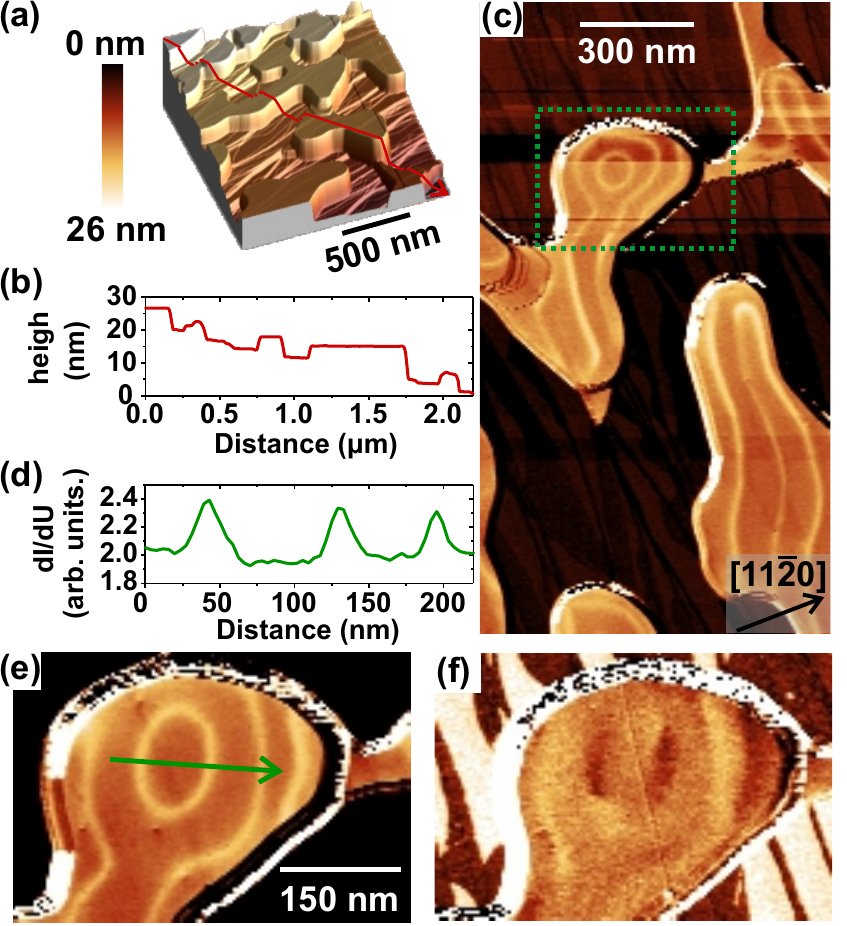}
  \end{center}
  \caption{(a) Large scale topographic STM image of 10 ML of Co
    deposited on Ru(0001) post annealed at $\approx450^\circ$C
    ($I$=\unit{1}{\nano\ampere}, $U$=\unit{-330}{\milli\volt}). (b)
    Line scan across the structure showing flat island surfaces and
    the general tilt of the Ru substrate. The typical local Co
    thickness is about 30 ML. (c)~Map of the $dI/dU$ signal showing a
    domain wall contrast ($I$=\unit{1}{\nano\ampere},
    $U$=\unit{-330}{\milli\volt},
    $\Delta{U}_{rms}$=\unit{30}{\milli\volt}). Zoomed image (e) and
    line scan (d) of the area marked by a green doted box of c). (f)
    Spin-polarized $dI/dU$ map recorded with a Co tip of in-plane
    spin-polarization ($I$=\unit{1}{\nano\ampere},
    $U$=\unit{-520}{\milli\volt},
    $\Delta{U}_{rms}$=\unit{50}{\milli\volt}). }
  \label{fig1}
\end{figure}

Co films were grown by molecular beam epitaxy from high purity Co rods
onto clean Ru(0001) surfaces. Details on the substrate preparation can
be found elsewhere \cite{Herve2017}. After deposition of 10 monolayers
(ML), the sample was annealed to $\approx$ 450$^\circ$C, which leads
to partial de-wetting of the film and the formation of flat islands
typically about 30 ML local thickness \cite{Yu2001}. Fig.~\ref{fig1}a
shows a typical topographic STM image of the surface. The line scan
(Fig.~\ref{fig1}b) displays a general tilt of the surface due to an
unavoidable miscut of the Ru substrate. The Co islands, however,
locally form atomically flat terraces. As it has been reported before,
Co grows in its hcp modification on Ru(0001) \cite{ElGabaly2006}.

Fig.~\ref{fig1}c shows a large area map of the differential
conductance $dI/dU$ of the sample taken with a non-magnetic W tip at a
bias voltage $U=$\unit{-330}{\milli\volt}. A strong contrast is found
between the thick hcp Co islands (bright) and the remaining thin
wetting layer (dark) due to large difference of their electronic
structure. Further a clear contrast can be resolved on the Co islands.
On the islands, white lines are found that either are closed loops or
end at the edges of the islands. Applying an out-of-plane magnetic
field leads to a movement of these white lines (see Supplementary)
identifying them as magnetic domain walls. Cobalt in its hcp
modification displays a strong uniaxial magnetic anisotropy of about  60\% of the
dipolar energy with an
easy axis along the c-axis of the hcp cell \cite{Hubert}. Thus, the Co film in this thickness range forms a
magnetic stipe domain pattern \cite{Hubert}, in which the out-of-plane
magnetocrystalline anisotropy orients the local magnetization normal
to the surface plane, i.e. the magnetization points out-of or into the
plane of the surface. The islands split into magnetic domains in the
form of a stripe domain pattern in order to reduce the dipolar energy
\cite{Hubert}.  Fig.~\ref{fig1}d shows a line scan across a
zoomed part of the sample displayed in Fig. \ref{fig1}e. As can be seen, the $dI/dU$ signal on
neighboring domains is identical and only the domain walls appear as
bright. This excludes that the observed signal is caused by the
tunneling magnetoresistance effect, in agreement with the non-magnetic
tip. Thus, the signal does not depend on the sign of the magnetization
but only on its orientation, i.e. out-of-plane (dark) on the domains
and in-plane (bright) on the domain walls. It thus is in accord with a
TAMR signal.  When dipping the W tip into a thick Co island, magnetic
Co can be transferred to the tip and the differential conductance then
becomes sensitive to the relative orientation of the local sample
magnetization and the tip magnetization due to the tunneling
magnetoresistance (TMR) effect
\cite{Juliere75,WiesendangerRevModPhys}.  As illustrated in
Fig.~\ref{fig1}f, this changes the observed contrast considerably. In
this case, the tip became sensitive to an in-plane component of the
sample magnetization and the signal of the domain walls depends on the
in-plane direction of the local magnetization.  They either appear as
bright or dark lines, depending on the direction of magnetization in
the walls. Most likely, the walls are of Bloch type due to the large
Co thickness \cite{Hubert}.

\begin{figure}[b]
  \includegraphics[width=8.2cm]{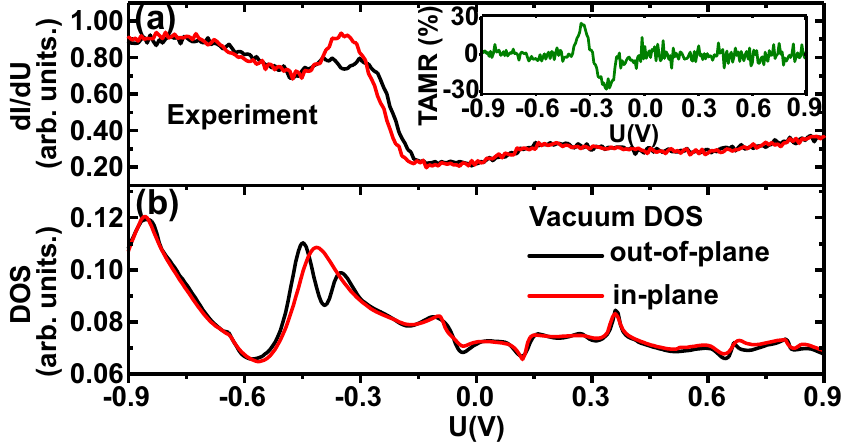}
  \caption{(a) Voltage dependence of the $dI/dU$ signal recorded with
    an unpolarized tip on in-plane magnetized domain walls (red) and
    on out-of-plane magnetized domains (black) displaying a sizable
    TAMR at $\approx$\unit{-350}{\milli\volt} (the tip was stabilized
    at $I$=\unit{1}{\nano\ampere}, $U$=\unit{-1}{\volt},
    $\Delta{U}_{rms}$=\unit{20}{\milli\volt}). The insert shows a TAMR of up to $\pm$ 30\%. (b)
    Calculated DOS in the vacuum 0.3 nm in front of the surface. The
    color scheme is the same as for a). While for most of the
    voltages, a sizable TAMR is absent, a characteristic dependence of
    the DOS is found near \unit{-350}{\milli\volt} in agreement with
    experiment.}
  \label{fig2}
\end{figure}

\begin{figure*}[t!]
  \includegraphics[width=17.8cm]{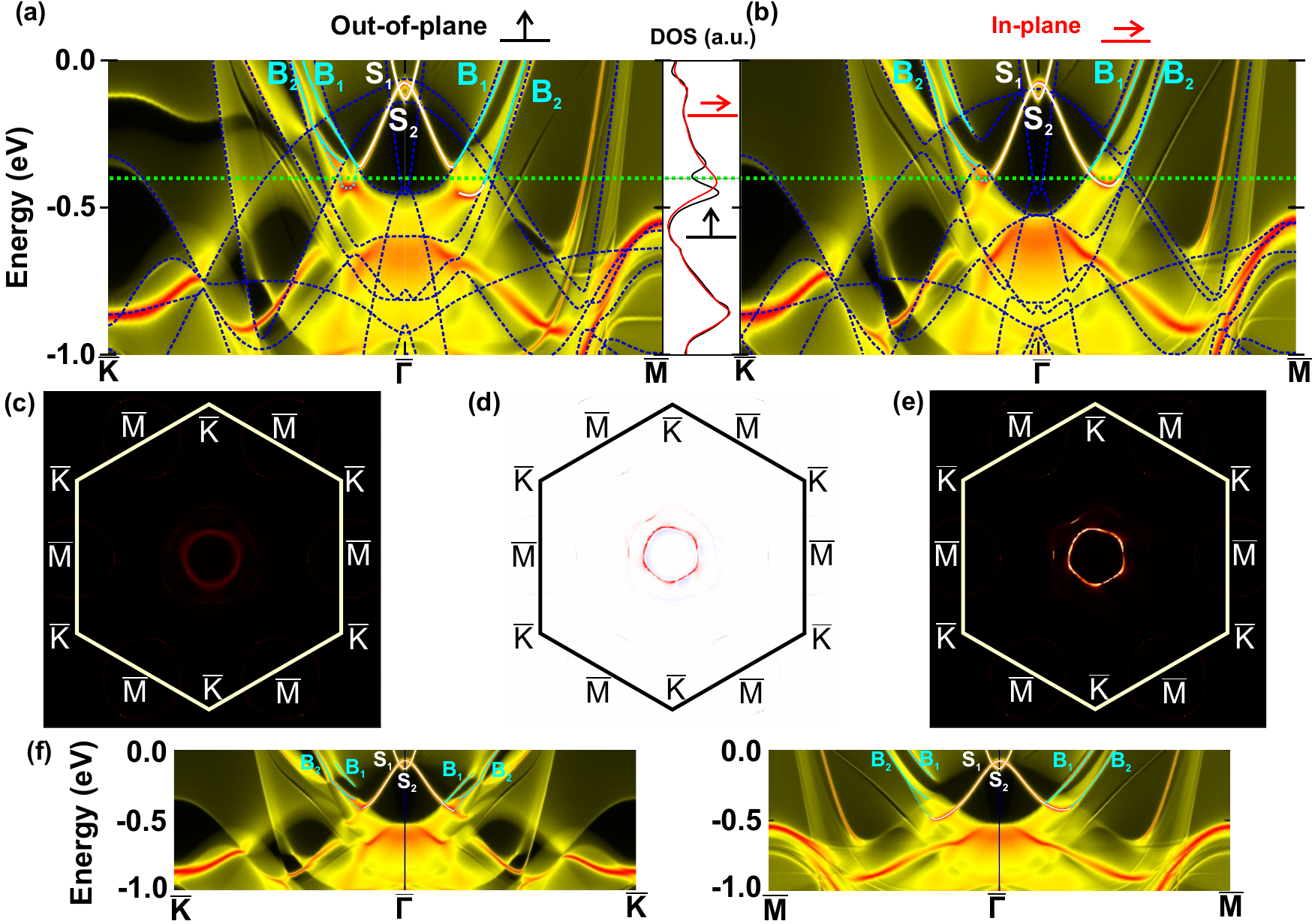}
  \caption{(a,b) Band structure of bulk Co (dashed dark blue lines) and
    the surface (yellow/red) for out-of-plane and in-plane magnetization,
    respectively. The surface and surface/bulk hybridized state are marked by white and light blue lines, respectively. For comparison the DOS of Fig.~\ref{fig2} has been
    repeated between the panels. (c,e) $k$-resolved vacuum DOS at \unit{-400}{\milli\volt} in the 2D surface Brillouin
    zone for out-of plane and in-plane magnetization,
    respectively using the same color scale. (d) Changes in the k-resolved density of
    states upon rotation of the magnetization. (f) Dispersion of the
    surface states for an in-plane magnetization along inequivalent
    directions in the surface Brillouin zone.}
  \label{fig3}
\end{figure*}

Since the TAMR effect is evoked by spin-orbit interaction causing
modifications in the DOS upon changes in the
magnetization axis, it is usually very dependent on the bias
voltage. In order to study the bias dependence, we recorded the
differential conductance $dI/dU$ on stripe domains with magnetization
out-of the plane and domain walls with an in-plane magnetization as
function of the bias voltage.  Fig.~\ref{fig2}a shows the result in a
wide voltage range between -900 and \unit{900}{\milli\volt}. At most
bias voltages, $dI/dU$ is within the noise identical on the
differently oriented magnetic structures.  Clear differences only
appear near \unit{-350}{\milli\volt}, where on the domain walls, a
broad peak is observed (red curve) while on the domains, a double peak
structure with a local minimum at that voltage is seen (black
curve). Note that at this bias voltage, a van Hofe singularity of an
occupied minority surface state of Co has been reported for bulk Co
and thin Co films
\cite{Himpsel1979,Diekhoener2003,Barral2005,Heinrich2010}. The inset
of the Figure shows the TAMR, i.e. the difference of the two
$dI/dU$ signals over the out-of plane signal (green curve), which quantifies the
TAMR effect. It clearly shows a resonance behavior around
\unit{-350}{\milli\volt} with a dip slightly below
\unit{-350}{\milli\volt} and a peak, above. The TAMR is surprisingly
large with up to 30\%. Note that a similar TAMR is even seen on a
single ML of Co on Ru(0001), but with a lower amplitude
\cite{Herve2017}.

In order to understand the origin of the TAMR, we carried out
first-principles calculations using a full potential relativistic
Green function method, specially designed for semi-infinite systems
such as surfaces and interfaces~\cite{Luders2001,Geilhufe2015}. The
calculations were performed within the density functional theory in a
generalized gradient approximation~\cite{Perdew1996}. $dI/dU$ signals
were simulated within the Tersoff-Hamann approximation, in which a
$dI/dU$ signal is associated with the local density of states (LDOS)
calculated in vacuum at a certain distance from the
surface~\cite{Tersoff1985}. In our study we calculated the LDOS at
\unit{3.5}{\AA} from the surface.

The calculated LDOS (Fig.~\ref{fig2}b) reproduces the measured $dI/dU$ signal
qualitatively. The overall shape of the LDOS agrees with the
experimental data. Especially, it shows nearly identical LDOS for both
magnetic configurations at most of the energies, i.e. an absence of a
TAMR effect. Only near \unit{-350}{\milli\volt}, the LDOS significantly
depends on the orientation axis of the magnetization. The calculated
LDOS nicely reflects the single peak structure for in-plane
magnetization (red) and the double peak structure for out-of plane
magnetization (black). Note that the calculations show a slightly
larger difference than observed in the experiment and also the
energies are slightly different. These small deviations from the
experiment may either be due to the limits of DFT or due to the
difference of the samples. In the experiments, we deal with Co films
of final thickness on Ru(0001), which may display some strain due to
the lattice mismatch to the substrate, while the calculations were
carried out for a half-infinite Co structure with its natural lattice
constant.

Finally, the DFT calculations allow to identify the origin of the
observed large TAMR effect. Fig.~\ref{fig3}a and c display the
calculated two-dimensional band structure at the Co surface. Bulk
states are superposed as dark blue dashed lines while states only present
at the surface are displayed in yellow/red. The intensity in the figure represents the weight of the states in a log-scale. For clarity, the surface and surface/bulk hybridized states discussed in the following are marked by white and light blue line respectively. For comparison, the LDOS of
Fig.~\ref{fig2}b is replotted vertically to size next to the band
structure. As can be seen, the difference in LDOS around
\unit{-350}{\milli\volt} are caused by large changes in the surface
band structure and are related to a forbidden band crossing. In the
following, we analyze the band structure in more detail.  As can be
seen from Fig. \ref{fig3}a) and b), the unoccupied surface state of
positive effective mass $S_1$ near the $\overline{\Gamma}$-point is
not affected by changes of the magnetization axis and is not involved
in the TAMR. Several of the bulk bands, which cross in case of an
out-of-plane magnetization, develop forbidden crossings upon rotating
the magnetization into the plane. This, however, hardly affects
the LDOS of the bulk (see Supplementary). The most prominent changes
in the surface electronic structure arises from changes in the surface
state $S_2$. The $S_2$ surface state of negative effective mass is of minority and $d_{3z^2-r^2}$ character
\cite{Wiebe2004,Barral2005,Heinrich2010}. For an in-plane
magnetization, $S_2$ merges
continuously with the bulk sp-band of positive effective mass $B_2$ causing
states with vanishing group velocity and a van Hofe singularity near 
-400 meV. Figure \ref{fig3}e displays a two-dimensional plot of the
states in the surface Brillouin zone and shows a bright ring at that energy as a
consequence (plotted on a linear scale). For an out-of-plane magnetization, $S_2$  hybridizes with $B_1$ and develops a gap
near -400meV reducing the density of states in the two-dimensional
Brillouin zone at that energy (compare Fig. \ref{fig3}c). This gap is responsible for
the reduction of the DOS at -400 meV and causes the TAMR. Its spectral weight is shifted away from the gap and causes the double peak feature in the DOS. This is further
illustrated by the change of the DOS depicted in Fig.\ref{fig3}d plotted in a linear scale (red for reduction an blue for an increase).  For
clarity, also cuts through the band structure along non-equivalent
directions are shown in Fig. \ref{fig3}f  for an in-plane
magnetization. While for an out-of-plane magnetization, the states at
$k$ and $-k$ are equivalent ($\vec{s}\perp\vec{k}$), for an in-plane
magnetization the bands experience a slight shift due to the
spin-momentum locking ($\vec{s}\vec{k}$) and direction dependent
hybridization with the bulk sp-bands.


In conclusion, we have shown that simple hcp Co shows a large TAMR of
$\pm$30\% near \unit{-350}{\milli\volt} caused by surface states.
Thus, this effect is expected to be rather robust to the details of
the Co film. Note that similar surface states have been reported even
for 2 ML Co films on Cu(111) and Au(111) of both fcc and hcp stacking
\cite{Heinrich2010} and Co on W(110) \cite{Wiebe2004}. As shown
recently, a similar contrast was even found for a single ML of Co on
Ru(0001) \cite{Herve2017}, further emphasizing the robustness of this
effect.

We acknowledge funding by the Deutsche Forschungsgemeinschaft (DFG) under the grant WU349/15-1, by the European Commission (Grant ATOMS FP7/2007-2013-62260), discussions with B. Dup\'e and M. Martins, and technical support from J. Chen, M. Peter and J. Jandke.

\bibliographystyle{apsrev}
\bibliography{./surfTAMR} 
\end{document}